\documentclass[preprint,showpacs,preprintnumbers,amsmath,amssymb]{revtex4}

\usepackage{graphicx}
\usepackage{dcolumn}
\usepackage{bm}
\usepackage{color}
\usepackage{booktabs}
\usepackage{multirow}

\usepackage{threeparttable}

\usepackage{epstopdf}

\usepackage[version=4]{mhchem}

\AtBeginDocument{
\heavyrulewidth=.08em
\lightrulewidth=.05em
\cmidrulewidth=.03em
\belowrulesep=.65ex
\belowbottomsep=0pt
\aboverulesep=.4ex
\abovetopsep=0pt
\cmidrulesep=\doublerulesep
\cmidrulekern=.5em
\defaultaddspace=.5em
}

\usepackage{hyperref}

\begin{document}


\title{Preparation of individual magnetic sub-levels of $^4$He($2^3$S$_1$) in a supersonic beam using laser optical pumping and magnetic hexapole focusing}

\author{Tobias Sixt}
\affiliation{Institute of Physics, University of Freiburg, Hermann-Herder-Str. 3, 79104 Freiburg, Germany}
\author{Jiwen Guan}
\altaffiliation{Current address: National Synchrotron Radiation Laboratory, University of Science and Technology of China, 230029, Hefei, P. R. China}
\affiliation{Institute of Physics, University of Freiburg, Hermann-Herder-Str. 3, 79104 Freiburg, Germany}
\author{Alexandra Tsoukala}
\affiliation{Institute of Physics, University of Freiburg, Hermann-Herder-Str. 3, 79104 Freiburg, Germany}
\author{Simon Hofs{\"a}ss}
\altaffiliation{Current address: Fritz-Haber-Institut der Max-Planck-Gesellschaft, Faradayweg 4-6, 14195 Berlin, Germany}
\affiliation{Institute of Physics, University of Freiburg, Hermann-Herder-Str. 3, 79104 Freiburg, Germany}
\author{Thilina Muthu-Arachchige}
\altaffiliation{Current address: Institute of Applied Physics, University of Bonn, Wegelerstr. 8, 53115 Bonn, Germany}
\affiliation{Institute of Physics, University of Freiburg, Hermann-Herder-Str. 3, 79104 Freiburg, Germany}
\author{Frank Stienkemeier}
\affiliation{Institute of Physics, University of Freiburg, Hermann-Herder-Str. 3, 79104 Freiburg, Germany}
\author{Katrin Dulitz\footnote{The author to whom correspondence may be addressed: katrin.dulitz@physik.uni-freiburg.de}}
\affiliation{Institute of Physics, University of Freiburg, Hermann-Herder-Str. 3, 79104 Freiburg, Germany}

\date{\today}

\begin{abstract}
\noindent We compare two different experimental techniques for the magnetic-sub-level preparation of metastable $^4$He in the $2^3$S$_1$ level in a supersonic beam, namely magnetic hexapole focusing and optical pumping by laser radiation. At a beam velocity of $v = 830\,$m/s, we deduce from a comparison with a particle trajectory simulation that up to $99\,$\% of the metastable atoms are in the $M_{J^{''}} = +1$ sub-level after magnetic hexapole focusing. Using laser optical pumping via the $2^3$P$_2-2^3$S$_1$ transition, we achieve a maximum efficiency of $94\pm3\,$\% for the population of the $M_{J^{''}} = +1$ sub-level. For the first time, we show that laser optical pumping via the $2^3$P$_1-2^3$S$_1$ transition can be used to selectively populate each of the three $M_{J^{''}}$ sub-levels ($M_{J^{''}} = $ -1, 0, +1). We also find that laser optical pumping leads to higher absolute atom numbers in specific $M_{J^{''}}$ sub-levels than magnetic hexapole focusing. 
\end{abstract}

\pacs{Valid PACS appear here}
\maketitle

\section{\label{sec:Intro}Introduction}
\noindent In a He gas discharge, two long-lived, excited (``metastable'') atomic levels are formed by electron-impact excitation from the $1^1$S$_0$ electronic ground level: the $2^3$S$_1$ level (electronic energy $E = 19.8\,$eV \cite{NIST_ASD}, natural lifetime $\tau = 7870\,$s \cite{Hodgman2009}) and the $2^1$S$_0$ level ($E = 20.6\,$eV \cite{NIST_ASD}, $\tau  = 19.7\,$ms \cite{VanDyck1971}). In the following, the metastable He atoms are referred to as spin-polarized, when only a single magnetic sub-level of He($2^3$S$_1$) is populated. Such spin-polarized metastable He (He$^{\mathrm{SP}}$) is used for a wide range of applications. Special interest is currently devoted to He magnetometry for the quantum sensing of very small magnetic fields, e.g. see Refs.\ \cite{Heil2017, Wang2020}. In metastable atom electron spectroscopy \cite{Harada1997}, also referred to as metastable de-excitation spectroscopy, He$^{\mathrm{SP}}$ has, for example, been used for probing surface magnetism \cite{Onellion1984}. In atom optics, He$^{\mathrm{SP}}$ has found applications in nanolithography, as well as in atomic waveguides and beamsplitters for atom interferometry \cite{Baldwin2005, Vassen2012}. He$^{\mathrm{SP}}$ also serves as a source of polarized electrons \cite{McCusker1969,McCusker1972} and ions \cite{Schearer1969}, e.g., for atomic and high-energy nuclear scattering experiments. Besides that, spin-polarized samples of $^3$He($2^3$S$_1$) are used for biomedical imaging, e.g., to visualize the human lung \cite{Walker1997, Kauczor1998, Nikolaou2015}.

Supersonic beams of He$^{\mathrm{SP}}$ are typically produced by optical pumping \cite{Happer1972}, as well as by magnetic (de-)focusing and magnetic deflection. Optical pumping of $^4$He($2^3$S$_1$) via the $2^3$P$_{J^{'}}-2^3$S$_1$ transition (where $J^{'} = 0, 1, 2$) at a wavelength of  $\lambda = 1083\,$nm has first been achieved by Franken, Colegrove and Schaerer using a helium lamp \cite{Franken1958, Colegrove1960, Schearer1961}. A few years later, also the optical pumping of the $2^3$S$_1$ level of the $^3$He isotope has been demonstrated using a similar setup \cite{Phillips1962, Colegrove1963}. The more recent use of narrowband laser radiation has proven to be particularly efficient for the optical pumping of He$^{\mathrm{SP}}$ \cite{Giberson1982, Lynn1990, Wallace1995, Granitza1995}.

Apart from that, also the interaction of a spin with an inhomogeneous magnetic field has been used for producing spin-polarized atomic beams of $^3$He($2^3$S$_1$) and $^4$He($2^3$S$_1$), respectively. These level-preparation techniques include Stern-Gerlach deflection \cite{Kato2012,Rubin2004,Zheng2017,Zheng2019,Chen2020,Smiciklas2010} , magnetic hexapole focusing \cite{Jardine2001, Woestenenk2001, Watanabe2006, Chaustowski2007,Kurahashi2008,Kurahashi2021,Baum1988} and Zeeman deceleration \cite{Dulitz2015a, Cremers2017b}.



A comparison between the different techniques for He$^{\mathrm{SP}}$ preparation in a supersonic beam is of paramount importance for experimental design considerations. In this article, we describe the results of a comparative study aimed at the laser optical pumping of $^4$He($2^3$S$_1$) into a single $M_{J^{''}}$ sub-level (where $M_{J^{''}} = -1, 0, +1$) and at the magnetic hexapole focusing (defocusing) of the $M_{J^{''}} = +1$ ($M_{J^{''}} = -1$) sub-level of $^4$He($2^3$S$_1$) using an array of two magnetic hexapoles. We have determined the efficiency for $M_{J^{''}}$-sub-level selection using low-cost fluorescence and surface-ionization detectors, respectively, which can easily be implemented in other experiments.

\section{\label{sec:Setup}Experiments}
\noindent Major parts of the experimental setup have already been described elsewhere \cite{Grzesiak2019,Guan2019}. Briefly, a pulsed $^4$He beam is produced by a supersonic expansion of $^4$He gas from a high-pressure reservoir ($30-40\,$bar) into the vacuum using a home-built CRUCS valve \cite{Grzesiak2018} ($30\,\mu$s pulse duration). An electron-seeded plate discharge (attached to the front plate of the valve) is used to excite an $\approx 4\cdot10^{-5}$ fraction of He atoms from the $1^1$S$_0$ electronic ground level to the two metastable levels, 2$^1$S$_0$ and 2$^3$S$_1$, referred to as He$^*$ hereafter \cite{Grzesiak2019}.
After passing through an $1\,$mm-diameter skimmer at a distance of $130\,$mm from the valve exit, the supersonic beam enters a second vacuum chamber, in which a specific magnetic sub-level of the 2$^3$S$_1$ level is prepared using laser optical pumping or selected using magnetic hexapole focusing (see below). The distance between the skimmer tip and the center of the optical pumping region (hexapole magnets) is $228\,$mm ($331\,$mm). 
The He$^*$ flux and the He$^*$ beam velocity are determined using Faraday cup detection at well-known positions along the supersonic beam axis $y$.
 
For the experiments on optical pumping, the pulsed valve is operated at room temperature resulting in a supersonic beam of He$^*$ with a mean longitudinal velocity of $v = (1844 \pm 6)\,$m/s ($250\,$m/s full width at half maximum, FWHM). For the experiments on magnetic hexapole focusing, the pulsed valve is cooled by a cryocooler (CTI Cryogenics, 350CP), and the valve temperature is actively stabilized to $42\,$K using PID-controlled resistive heating. This results in a supersonic beam of He$^*$ with a mean longitudinal velocity of $v = (830 \pm 17)\,$m/s ($\approx 130\,$m/s FWHM).

\subsection{\label{sec:setup.optPump} Laser optical pumping}
\begin{figure}[hbt!]
	\includegraphics{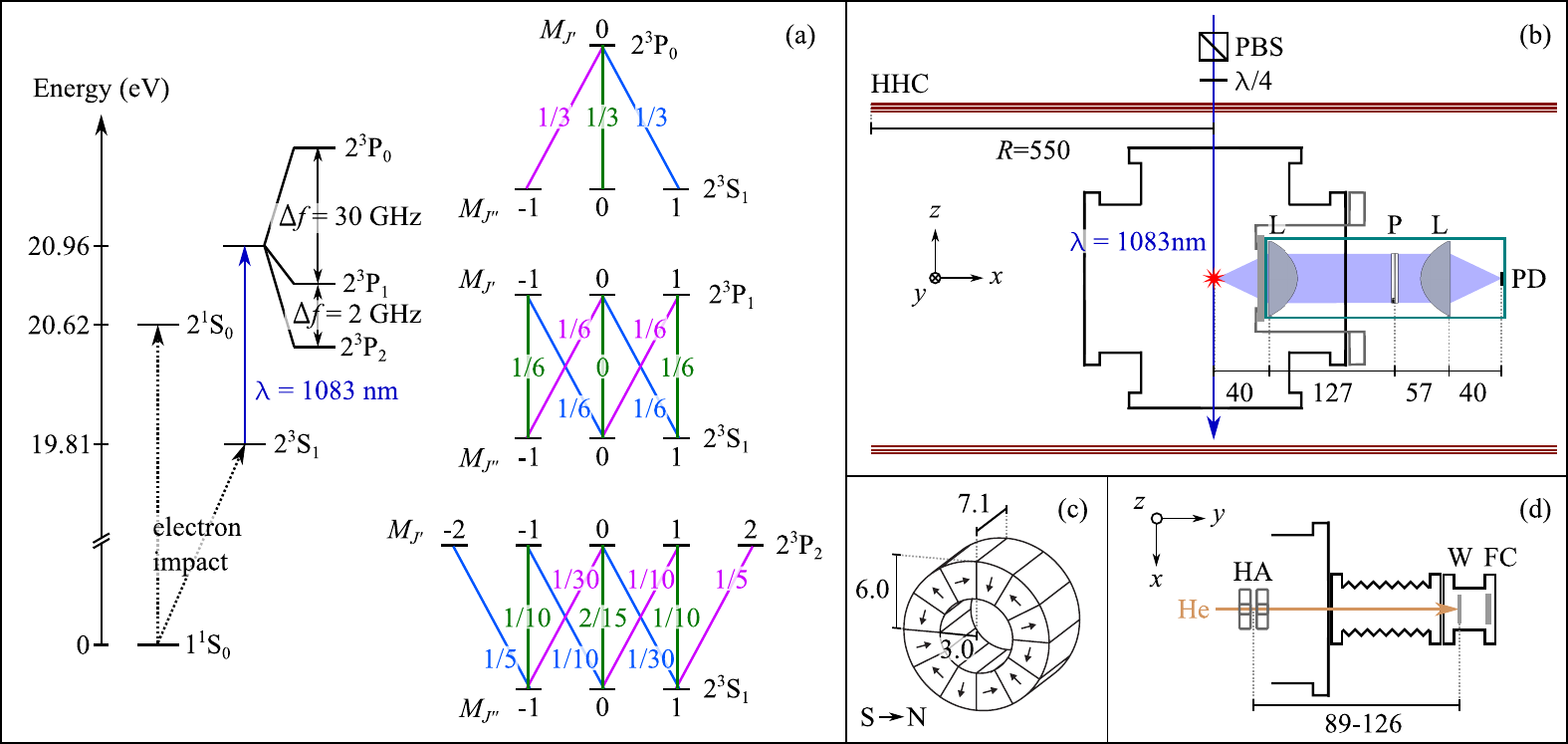}
	\caption{\label{fig:Setup}	
	(a) Left: He energy levels relevant for the experiments described in the main text. The level energies are taken from Ref.\ \cite{NIST_ASD}. Right: Transitions relevant for the laser optical pumping of He($2^3$S$_1$) via the $2^3$P$_0$ (top), the $2^3$P$_1$ (middle) and the $2^3$P$_2$ (bottom) levels, respectively. The relative transition strengths for $\sigma^+$, $\pi$ and $\sigma^-$ excitation are labeled in pink, green and blue color, respectively.
	(b) Schematic illustration of the experimental setup used for optical pumping and fluorescence detection. HHC = Helmholtz coil, PBS = polarizing beam splitter, $\lambda/4$ = quarter wave plate, PD = photodiode, P = polarizer, L = aspheric lens.
	(c) Schematic drawing of a magnetic hexapole array in Halbach configuration.
	(d) Sketch of the detection system for magnetic hexapole focusing including the two Halbach arrays (HA), the wire detector (W) and a Faraday cup detector (FC). All dimensions in (b)-(d) are given in units of mm, and they are not to scale.}
\end{figure}

\noindent The energy-level schemes and the experimental setup used for laser optical pumping are shown in Figs.\ \ref{fig:Setup} (a) and (b), respectively. Optical pumping is achieved by laser excitation via the $2^3$P$_1-2^3$S$_1$ transition or via the $2^3$P$_2-2^3$S$_1$ transition at $\lambda =$ 1083 nm, respectively. The laser light for optical pumping is generated by a combination of a fiber laser and a fiber amplifier (NKT Photonics, Koheras BOOSTIK Y10 PM FM, $2.2\,$W output power, $10\,$kHz line width). The laser frequency is stabilized using frequency-modulated, saturated absorption spectroscopy in a He gas discharge cell. Since the frequency difference between the $2^3$P$_1$ and $2^3$P$_2$ spin-orbit levels is only $\Delta f \approx 2\,$GHz \cite{NIST_ASD}, the laser frequency can be changed in between the different transitions without effort.

The laser light is guided to the vacuum chamber using a polarization-maintaining single-mode fiber, where it is collimated to a beam diameter of $2 w_0 \approx 14\,$mm ($w_0$ is the Gaussian beam waist). Before the laser beam enters the vacuum chamber, it passes a polarizing beam splitter for polarization clean-up, and a quarter wave plate for the production of circularly polarized light. Inside the vacuum chamber, the laser beam crosses the supersonic beam at right angles and parallel to the direction of the magnetic field produced by two coils in near-Helmholtz configuration (radius $R = 55\,$cm). The thus produced homogeneous magnetic field of $ B_z \leq 4.5\,\text{G}$ 
provides a uniform quantization axis for all He atoms in the supersonic beam.

The level-preparation efficiency is determined by measuring the laser induced-fluorescence (LIF) of the He atoms in the direction perpendicular to the supersonic beam and the laser beam. The fluorescence light is collected and focused onto an InGaAs photodiode (Hamamatsu, $1\,$mm active area diameter, photosensivity of $R_{\text{PD}} = 0.63\,$A/W at $\lambda = 1080\,\text{nm}$) using two anti-reflection-coated, aspheric lenses (Thorlabs, $75\,$mm diameter, $60\,$mm focal length). Due to the symmetric lens configuration (as shown in Fig.\ \ref{fig:Setup} (b)), the fluorescence collection region in the $yz$ plane is expected to be of the same size as the detection region, which is given by the active area of the photodiode. The output of the photodiode is amplified using a home-built transimpedance amplifier with a gain of $G_{\text{PD}} \approx 5 \cdot 10^{5}\,\text{V}/\text{A}$. A rotatable linear polarizer (Thorlabs, extinction ratio $> 400:1$ at $\lambda = 1083\,\text{nm}$) is mounted in between the lenses in order to analyze the polarization of the fluorescence light. All the optical components of the fluorescence detector are placed into a single lens tube system to ensure the correct alignment of the optical parts. The entire detector assembly is positioned on a translational stage outside the vacuum chamber which can be moved along the $y$ axis.

Under normal operating conditions, the number of He atoms in the $2^3$S$_1$ level is $\approx 10^9$/pulse, as inferred from the signal on the Faraday-cup detector \cite{Grzesiak2019}. For excitation via the $2^3$P$_2-2^3$S$_1$ transition, the time-dependent signal of the photodiode has a peak voltage of $U_{\text{PD,max}} \approx 41\,\text{mV}$ and an FWHM of 27 $\mu$s. The peak flux of detected photons is then inferred from $U_{\text{PD,max}}$ using
\begin{align}
	\dot{N}_{\text{ph,max}} = \frac{U_{\text{PD,max}} }{h \nu R_\text{PD} G_{\text{PD}}} \approx 7 \cdot 10^{11}\,\frac{\text{photons}}{\text{s}},
\end{align}
where $h$ is Planck's constant and $\nu$ is the corresponding transition frequency. From these measurements, we infer a root-mean-square noise amplitude of $U_{\text{noise}} = 6.4\,\text{mV}$ for a single measurement which improves to $U_{\text{noise}} = 0.4\,\text{mV}$ by averaging over 300 gas pulses. This results in a signal-to-noise ratio of 
\begin{align}
	SNR = \frac{U_{\text{PD,max}}^2}{U_{\text{noise}}^2} =
	\begin{cases}
	16.2\,\text{dB} \quad \text{single measurement}\\
	40.3\,\text{dB} \quad \text{300 averages}
	\end{cases}
\end{align}
At $SNR = 0\,\text{dB}$, we thus infer a detection limit of $\dot{N}_{\text{ph,lim}} \approx 1 \cdot 10^{11}\,$photons/s ($\dot{N}_{\text{ph,lim}} \approx 7 \cdot 10^{9}\,$photons/s) for a single measurement (300 averages).


\subsection{\label{sec:setup.magDefl} Magnetic hexapole focusing}
\noindent For the magnetic focusing of He(2$^3$S$_1$, $M_{J^{''}} = +1$), we use a set of two Halbach arrays \cite{Halbach1980, Halbach1981} in hexapole configuration, sketched in Fig.\ \ref{fig:Setup} (c), whose design has already been described previously \cite{Dulitz2014a, Dulitz2016}. Each hexapole array consists of 12 magnetized segments (Arnold Magnetic Technologies, NdFeB, N42SH, remanence of $B_0 = 1.3\,$T) which are glued into an aluminium housing and placed on a position-adjustable rail at a center-to-center distance of $14.6\,$mm.

To determine the focusing properties of the magnet assembly, a thin stainless-steel wire (diameter $d_{\mathrm{wire}} = 0.2\,$mm, labelled as ``W'' in Fig.\ \ref{fig:Setup} (d)) is used as a position-sensitive Faraday-cup-type detector. Its position along the $y$ and $x$ axes can be varied by a maximum of $180\,$mm and $50\,$mm, respectively, using a set of two precision linear translators. A second Faraday-cup detector (labelled as ``FC'' in Fig.\ \ref{fig:Setup} (d)), i.e., a stainless-steel plate of $30\,$mm diameter, is placed behind the wire detector to determine the He$^*$ beam velocity. 

\section{Results and Discussion}
\subsection{\label{sec:optPump} Laser optical pumping}
\noindent In general, the level preparation efficiency relies on the polarization state of the laser radiation, on the energy-level structure of the involved electronic levels and on the transition strengths for photon absorption and emission. If atoms are excited with $\sigma^{+ (-)}$-polarized light, the change in angular momentum between the upper and lower level is $\Delta M_{J^{''}} = M_{J^{'}}-M_{J^{''}} = +1\,(-1)$ for every photon-scattering event, where $M_{J^{'}}$ and $M_{J^{''}}$ are the magnetic projection quantum numbers for the upper and the lower magnetic sub-levels, respectively. For excitation with $\pi$-polarized light, $\Delta M_{J^{''}} = 0$.

The transition strengths for the $2^3$P$_{J^{'}}-2^3$S$_1$ transitions (where $J^{'} = 0, 1, 2$) in He are shown in Fig.\ \ref{fig:Setup} (a). As can be seen from the figure, all $M_{J^{'}}- M_{J^{''}}$ transitions strengths for the $2^3$P$_2-2^3$S$_1$ transition are non-zero. Hence, in this case, multiple excitation cycles with $\sigma^{+ (-)}$-polarized light lead to equal populations in the 2$^3$S$_1$, $M_{J^{''}} = +1\,(-1)$ and 2$^3$P$_2$, $M_{J^{'}} = +2\,(-2)$ sub-levels, respectively. In this case, photon emission via this transition continues to occur as long as the atoms are subject to laser excitation. In contrast to that, the emission of photons ceases after a few pumping cycles for $\sigma^{+(-)}$ excitation of the $2^3$P$_1-2^3$S$_1$ transition, since all population is trapped in the 2$^3$S$_1$, $M_{J^{''}} = +1\,(-1)$ sub-level. Likewise, the excitation of the $2^3$P$_1-2^3$S$_1$ transition using $\pi$-polarized light leads to a transfer of population into the $2^3$S$_1, M_{J^{''}} = 0$ sub-level and photon emission stops as a result of the zero transition strength for the $2^3$P$_1, M_{J^{'}}=0 - 2^3$S$_1, M_{J^{''}}=0$ transition.

As can be inferred from Fig.\ \ref{fig:Setup} (a), the selective population of a single $M_{J^{''}}$ sub-level in He($2^3$S$_1$) via the $2^3$P$_0-2^3$S$_1$ transition is more complicated, as it requires two different laser polarization states. Therefore, we have focused our experimental work on the optical pumping of He($2^3$S$_1$) via the $2^3$P$_2-2^3$S$_1$ and $2^3$P$_1-2^3$S$_1$ transitions, respectively. In the following, we provide a brief description of the different optical pumping schemes and the methods used for analyzing and optimizing the sub-level preparation efficiencies. Furthermore, we compare our results with literature values.

\subsubsection{\label{sec:optPumpAnalysis3P2} Optical pumping via the $2^3$P$_2-2^3$S$_1$ transition}
\begin{figure}[hbt!]
	\includegraphics[width=8cm]{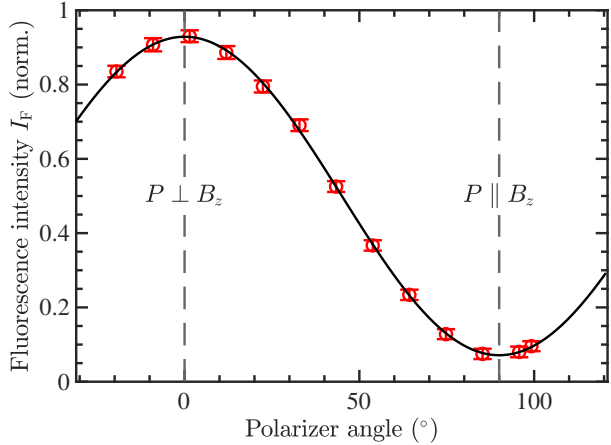}
		\caption{\label{fig:P2FLmodulation} Fluorescence intensity as a function of polarizer angle for excitation with $\sigma^+$-polarized light via the $2^3$P$_2-2^3$S$_1$ transition. The experimental data points are shown as red circles. The black curve is a sine fit to the data. The dashed vertical lines represent the two angles at which the transmission axis of the polarizer $P$ is perpendicular or parallel to the magnetic field component $B_z$, respectively. Here, an efficiency of $\eta_{+1} \approx 93\,$\% is determined from the fit to the data.}
\end{figure}

\noindent As stated above, the excitation of the $2^3$P$_2-2^3$S$_1$ transition with $\sigma^{+ (-)}$-polarized light leads (after a few excitation cycles) to the optical cycling between the $2^3$S$_1$, $M_{J^{''}} = +1\,(-1)$ sub-level and the 2$^3$P$_2$, $M_{J^{'}} = +2\,(-2)$ sub-level. In this case, the atomic fluorescence only consists of $\sigma^{+ (-)}$-polarized light as $\Delta M = M_{J^{'}} - M_{J^{''}} = +1\, (-1)$. During the first optical pumping cycles, and for a non-perfect circular polarization of the input light, also sub-levels with $M_{J^{'}} \neq +2\,(-2)$ are populated and can decay back to the $2^3$S$_1$ level while emitting $\sigma^{- (+)}$- and $\pi$-polarized photons as well. Therefore, the polarization purity of the emitted fluorescence provides information about the sub-level preparation efficiency. Since the polarization of fluorescence photons are given with respect to the quantization axis, which is the pointing of the external magnetic field vector, fluorescence photons of $\sigma^{+/-}$ polarization that are detected along an axis perpendicular to the quantization axis, are projected onto a linear polarization. The direction of this projected linear polarization is again perpendicular to the quantization axis. Furthermore, fluorescence photons of $\pi$ polarization are linearly polarized parallel to the quantization axis. Therefore, we analyze the purity of the emitted light using a linear polarizer plate and the fluorescence intensity can be ascribed to $\sigma^{+/-}$ polarization ($\pi$ polarization) if the transmission axis of the polarizer $P$ is perpendicular (parallel) to the magnetic field component $B_z$, respectively \cite{Hubele2015}. For example, the results of fluorescence measurements at different polarizer angles are shown in Fig.\ \ref{fig:P2FLmodulation}.

In order to compare the sub-level preparation efficiencies, we define
\begin{equation}
\eta_i = \frac{\rho(M_{J^{''}_i})}{\sum_{i}{\rho(M_{J^{''}_i})}},
\end{equation}
for producing a specific magnetic sub-level population $\rho(M_{J^{''}_i})$ of He(2$^3$S$_1$) (where $i = -1, 0, +1$). For the $2^3$P$_2-2^3$S$_1$ transition, the efficiency for optical pumping into the $2^3$S$_1, M_{J^{''}} = +1\,(-1)$ sub-level is thus obtained using
\begin{align}
\eta_{+1\,(- 1)} = 1 - \frac{I_{\mathrm{F}}(P \parallel B_z)}{I_{\mathrm{F}}(P \parallel B_z) + I_{\mathrm{F}}(P \perp B_z)},
\end{align}
where $I_{\mathrm{F}}(P \parallel B_z)$ and $I_{\mathrm{F}}(P \perp B_z)$ are the fluorescence intensities for emission at polarizer axes $P \parallel B_z$ and $P \perp B_z$, respectively.

\subsubsection{\label{sec:optPumpAnalysis3P1} Optical pumping via the $2^3$S$_1 - 2^3$P$_1$ transition}
\begin{figure}[hbt!]
	\includegraphics[width=8cm]{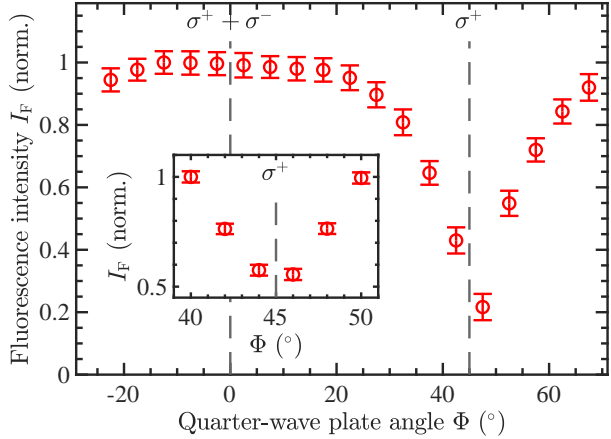}
	\caption{\label{fig:P1FLmodulation} Red circles: Measured fluorescence intensity for excitation via the $2^3$P$_1-2^3$S$_1$ transition as a function of quarter wave plate angle $\Phi$. The quarter wave plate angles for excitation with pure $\sigma^+$-polarized light and with an equal mixture of $\sigma^+$- and $\sigma^-$-polarized light are indicated as dashed vertical lines. The inset shows the results of a measurement in a region around $\Phi = 45^\circ$ taken under different experimental conditions.
	}
\end{figure}

\noindent Excitation via the $2^3$P$_1-2^3$S$_1$ transition allows for the selective optical pumping into each of the $M_{J^{''}}$ sub-levels in He($2^3$S$_1$). When the atoms are excited with pure $\sigma^{+(-)}$-polarized light, all population is pumped into the $2^3$S$_1, M_{J^{''}} = +1\,(-1)$ sub-level. Since this is a dark sub-level, fluorescence emission should only occur in the first few pumping cycles. However, by using a mixture of $\sigma^{+}$- and $\sigma^{-}$-polarized excitation light, the dark sub-level is remixed, so that optical cycling (and thus fluorescence emission) continues to occur. In the present configuration, the input polarization is changed by varying the angle $\Phi$ of the quarter wave plate compared to the axis of the incident linear laser polarization. The observed change of the fluorescence intensity as a function of quarter wave plate angle is shown in Fig.\ \ref{fig:P1FLmodulation}. The efficiency for pumping into the $2^3$S$_1, M_{J^{''}} = +1\,(-1)$ sub-level is determined using
\begin{align}
\eta_{+1\,(- 1)} = 1 - \frac{I_{\text{F}}(\sigma^{+(-)})}{I_{\text{F}}(\sigma^{+} + \sigma^{-})},
\end{align}
where $I_{\mathrm{F}}(\sigma^{+(-)})$ and $I_{\mathrm{F}}(\sigma^{+}+\sigma^{-})$ are the fluorescence intensities for excitation with pure $\sigma^{+(-)}$ polarization and with a mixture of $\sigma^{+}$ and $\sigma^{-}$ polarization, respectively.

\begin{figure}[hbt!]
	\includegraphics[width=8cm]{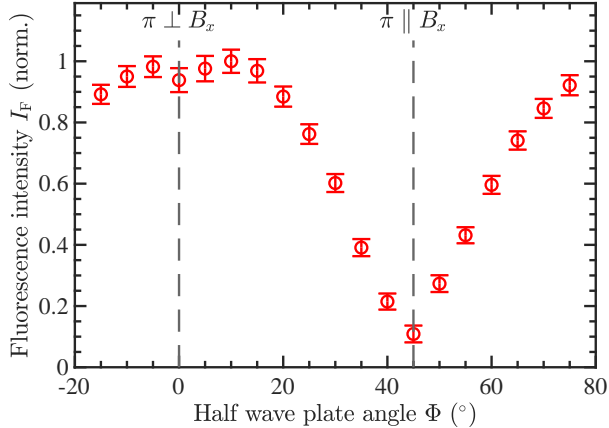}
	\caption{\label{fig:P1M0FLmodulation} Red circles: Measured fluorescence intensity for excitation via the $2^3$P$_1-2^3$S$_1$ transition as a function of half wave plate angle $\Phi$. The half wave plate angles for excitation with $\pi$-polarized light parallel and perpendicular to the magnetic field direction $B_x$ are indicated as dashed vertical lines. 
	}
\end{figure}

For optical pumping into the $2^3$S$_1(M_{J^{''}} = 0)$ sub-level, we have used an additional coil pair in near-Helmholtz configuration (radius of $76\,$mm, distance of $255\,$mm), placed at right angles to the other Helmholtz-coil pair, to generate a well-defined quantization axis along the $x$ direction. As a result, the laser beam direction is perpendicular to the magnetic field component $B_x$. In addition to that, the quarter wave plate is replaced by a half wave plate. By rotating the half wave plate, the angle of polarization is adjusted to be either parallel or perpendicular to $B_x$. In the latter case, the excitation light is projected onto an equal mixture of $\sigma^{+}$ and $\sigma^{-}$ input polarization which again causes a remixing of the otherwise dark sub-level $2^3$S$_1, M_{J^{''}} = 0$. Fig.\ \ref{fig:P1M0FLmodulation} shows the change of the fluorescence intensity as a function of half wave plate angle. For pumping into the $2^3$S$_1, M_{J^{''}} = 0$ sub-level, the sub-level preparation efficiency is thus obtained using
\begin{align}
\eta_{0} = 1 - \frac{I_{\text{F}}(\pi \parallel B_x)}{I_{\text{F}}(\pi \perp B_x)},
\end{align}
where $I_{\mathrm{F}}(\pi \parallel B_x)$ and $I_{\mathrm{F}}(\pi \perp B_x)$ are the fluorescence intensities for excitation using $\pi$-polarized light in a direction parallel and perpendicular to the magnetic field component $B_x$, respectively.

\subsubsection{\label{sec:optPumpPosition} Optimization of the sub-level preparation efficiency}
\noindent We have identified several parameters which strongly affect the sub-level preparation efficiency: the interaction time between the excitation laser light and the sample, the laser intensity, the magnetic field strength and the purity of the input polarization.

During the excitation process, an atom typically scatters several photons before it is pumped to the designated magnetic sub-level. Since the radiative lifetime of the 2$^3$P$_{J^{'}}$ levels in He is long compared to typical optical pumping transitions in other atoms ($\tau = 1/\Gamma = 97.89\,\text{ns}$ \cite{NIST_ASD}), a comparably long interaction time between the laser beam and the sample has to be achieved. In our case, we have found that a large 1/$e^2$ laser beam diameter of $2 w_0 \approx 14\,\text{mm}$ is most practical for this purpose. For a supersonic beam with a mean velocity of $1844\,$m/s, this beam diameter translates into an interaction time of $\Delta t_{\mathrm{int}} = 7.6\,\mu\text{s} \gg \tau$.

\begin{figure}[hbt!]
	\includegraphics[width=8cm]{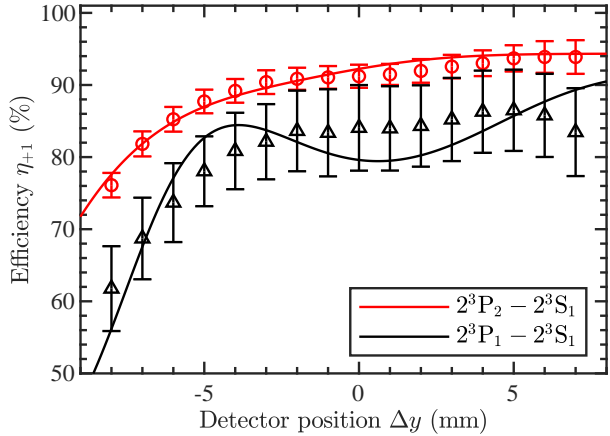}
	\caption{\label{fig:optPumpPosition} sub-level preparation efficiencies $\eta_{+1}$ for $\sigma^+$ excitation of the $2^3$P$_2-2^3$S$_1$ and $2^3$P$_1-2^3$S$_1$ transitions (see legend), respectively, at different positions of the fluorescence detector along the $y$ axis. The origin of the position axis denotes the midpoint of the laser beam.  Experimental values are shown as markers, and the results of the rate-equation calculations are shown as solid lines. In the calculations, a mixture of $95\,$\% $\sigma^+$- and $5\,$\% $\sigma^-$-polarized light is assumed for both excitation schemes.}
\end{figure}

We have studied the influence of the interaction time on the sub-level preparation efficiency $\eta_i$ by monitoring the fluorescence intensity at different fluorescence detector positions along the $y$ axis. As can be seen from the colored markers in Fig.\ \ref{fig:optPumpPosition}, the efficiency $\eta_{+1}$ for $\sigma^+$ excitation of the $2^3$P$_2-2^3$S$_1$ and $2^3$P$_1-2^3$S$_1$ transitions, respectively, increases to a nearly constant value as the detector is moved towards the midpoint of the excitation laser beam. This confirms that, in our experiment, the interaction time does not limit the sub-level preparation efficiency. 

We have simulated the population transfer process using rate-equation calculations. A detailed description of the rate-equation model can be found in App. \ref{app:ratEqns}. The best fit to our experimental data for excitation via the $2^3$P$_2-2^3$S$_1$ transition and via the $2^3$P$_1-2^3$S$_1$ transition, respectively, is found by assuming that the excitation light is a mixture of $95\,$\% $\sigma^+$- and $5\,$\% $\sigma^-$-polarized light. The admixture of wrongly polarized light also explains why the observed sub-level preparation efficiency is below $100\,\%$. In addition to that, as can be seen from Fig.\ \ref{fig:Setup} (a), the relative transition strengths for optical pumping with wrongly polarized light is $1/6$ for the $2^3$P$_1-2^3$S$_1$ transition, while it is only $1/30$ for the $2^3$P$_2-2^3$S$_1$ transition. Thus, optical pumping via the $2^3$P$_1-2^3$S$_1$ transition is more sensitive to wrongly polarized excitation light which explains the observed difference in the sub-level preparation efficiency. In our setup, such an admixture of wrong input polarization might be caused by imperfections of the quarter wave plate or by the birefringence of the vacuum window.

\begin{figure}[hbt!]
	\vspace{0.5cm}
	\includegraphics[width=8cm]{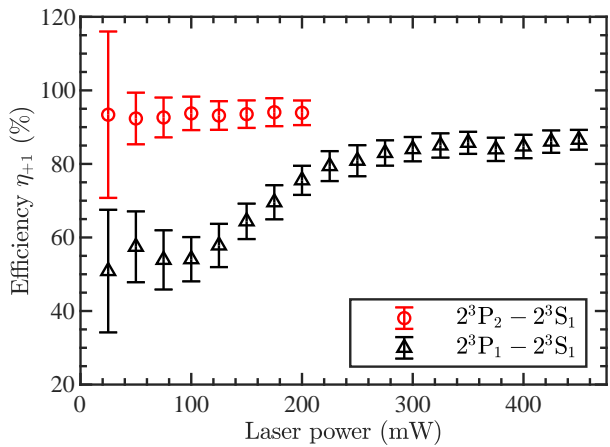}
	\caption{\label{fig:optPumpPower} sub-level preparation efficiency $\eta_{+1}$ for $\sigma^+$ excitation of the $2^3$P$_2-2^3$S$_1$ and $2^3$P$_1-2^3$S$_1$ transitions (see legend), respectively, at different laser powers. The data are taken at a $6\,$mm distance downstream from the midpoint of the laser beam in order to represent the efficiencies at equilibrium.}
\end{figure}

Secondly, the laser intensity has to be high enough so that the laser-induced power broadening compensates for a detuning of the laser frequency from the atomic resonance. This detuning is caused by the Doppler broadening due to the transverse velocity of the atoms ($\Delta_{\text{Doppler}} \approx 12\,\text{MHz}$ FWHM) and by the Zeeman shift of the atomic levels ($\Delta_{\mathrm{Zeeman}} < 14\,\text{MHz}$). The FWHM of the power broadening can be expressed as
\begin{align}
\Delta_{\text{power}} = \frac{\Gamma}{2 \pi} \cdot \sqrt{1 + \frac{I}{I_\text{sat}}},
\end{align}
where $I$ is the intensity of the laser light and $I_{\text{sat}} \approx 0.16\,$mW/cm$^2$ (assuming a two-level system) is the saturation intensity of the transition. Therefore, in order to compensate for the Doppler broadening and for the Zeeman shift, the laser intensity has to be $I \geq 12 \,\frac{\text{mW}}{\text{cm}^2}$, corresponding to a laser power of $\geq 9\,\text{mW}$ for our experiments. From Fig.\ \ref{fig:optPumpPower}, we can see that the sub-level preparation efficiency for $\sigma^+$ excitation of the $2^3$P$_2-2^3$S$_1$ transition is constant for laser powers $P > 50\,\text{mW}$. Unfortunately, measurements of the sub-level preparation efficiency at lower laser powers suffer from low signal intensities and are thus less representative. 
For $\sigma^+$ excitation of the $2^3$P$_1-2^3$S$_1$ transition, we observe that more than 300 mW of laser power are required to reach a constant sub-level preparation efficiency. 
This power difference might be attributed to a weaker power broadening of the $2^3$P$_1-2^3$S$_1$ line compared to the $2^3$P$_2-2^3$S$_1$ line as a result of a higher saturation intensity for this transition. As both transitions have the same line width, the same initial level and approximately the same transition frequency, we can see from Eq. \eqref{eq:dmd} that the squared dipole matrix elements $\left|\mu_{J^{'}}\right|^2$ are proportional to the degeneracy factors $2J^{'} + 1$. As $I_{\text{sat}} \propto 1/ \left| \mu_{J^{'}} \right|^2$, it follows that $I_{\text{sat}}(2^3$P$_1-2^3$S$_1) / I_{\text{sat}}(2^3$P$_2-2^3$S$_1) = \left|\mu_2\right|^2 / \left|\mu_1 \right|^2=  5/3$.

Thirdly, the magnetic bias field has to be large enough to ensure a uniform quantization axis within the optical pumping region so that the contributions of stray fields along other spatial directions is small. 

\begin{figure}[hbt!]
	\includegraphics[width=8cm]{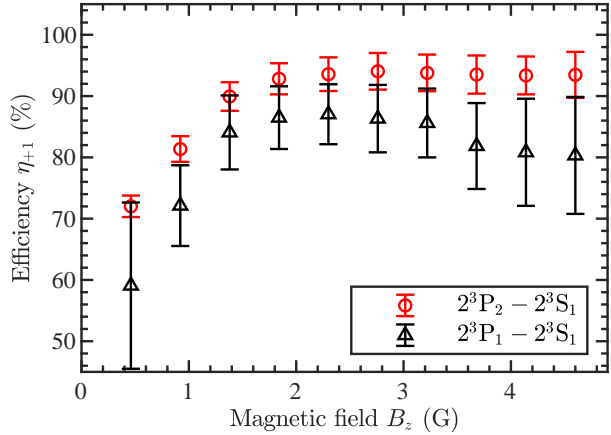}
	\caption{\label{fig:optPumpBfiled} sub-level preparation efficiency $\eta_{+1}$
		for $\sigma^+$ excitation of the $2^3$P$_2-2^3$S$_1$ and $2^3$P$_1-2^3$S$_1$ transitions (see legend), respectively, at different bias magnetic field strengths. The data are taken at a $6\,$mm distance downstream from the midpoint of the laser beam in order to represent the efficiencies at equilibrium.}
\end{figure}

In Fig.\ \ref{fig:optPumpBfiled}, a scan of the sub-level preparation efficiency $\eta_{+1}$ for $\sigma^+$ excitation of the $2^3$P$_2-2^3$S$_1$ and $2^3$P$_1-2^3$S$_1$ transitions, respectively, is shown as a function of the magnetic field component $B_z$. The highest efficiency is achieved at field strengths between $2\,\mathrm{G} \leq B_z \leq 3\,\mathrm{G}$ for both transitions. This magnetic field range is in line with previous observations reported in the literature \cite{Gillot2013,Wallace1995,Schearer1990a}. At magnetic field strengths $B_z > 3\,\mathrm{G}$, the sub-level preparation efficiency for excitation via the $2^3$P$_1-2^3$S$_1$ ($2^3$P$_2-2^3$S$_1$) transition is decreased (remains constant) compared to the optimum $B_z$ field range. This is consistent with a decreased scattering rate at higher magnetic fields caused by the increased Zeeman detuning. 

We have also analyzed the influence of stray magnetic fields along the $x$ and $y$ directions. Using a high-accuracy, three-axis Gauss probe (Stefan Mayer Instruments, $\leq 1\,\mathrm{G}$, $0.05\,$mG resolution), we obtain $B_x \approx 0.2\,\text{G}$ and $B_y \approx 0.1\,\text{G}$. At $B_z = 3\,\text{G}$, this results in an angle of $\theta = \sqrt{B_x^2+B_y^2}/B_z \approx 80\,\text{mrad}$ 
between the magnetic field and the $z$ axis (cf. Gillot et al. \cite{Gillot2013}). We have observed that a further compensation of the magnetic stray fields using additional coils along the $x$ axis (resulting in $\theta < 40\,\text{mrad}$) does not result in an improved sub-level preparation efficiency. In addition to that, a non-perfect alignment of the laser propagation direction parallel to the quantization axis can induce a similar limit to the achievable sub-level preparation efficiency as the presence of magnetic stray fields. Furthermore, small magnetic-field inhomogeneities within the interaction region, resulting from e.g. a not perfect Helmholtz coil arrangement or electronic devices in the laboratory, may also limit the sub-level-preparation efficiency.

In summary, we conclude that the imperfect polarization of the laser light (see discussion above) is the main limiting factor for the sub-level preparation efficiency.

\subsubsection{\label{sec:compOPlit} Comparison with literature values}
\noindent In Tab.\ \ref{tab:optPumpEfficiencies}, we present a summary of the maximum sub-level preparation efficiencies $\eta_{i,\mathrm{max}}$ obtained from our experimental data, and a comparison with literature values. As can be seen from the table, our $\eta_{i,\mathrm{max}}$ values are in good agreement with previous results for the laser optical pumping of He($2^3$S$_1$). To the best of our knowledge, we are the first to obtain a maximum efficiency $> 90\,$\% for optical pumping into $M_{J^{''}} = 0$. The only previous attempt to selectively populate $M_{J^{''}} = 0$ has been by Giberson et al. \cite{Giberson1982} using linearly polarized light resonant with the $2^3$P$_0-2^3$S$_1$ transition and propagating along the quantization axis.

For optical pumping into the spin-stretched sub-levels ($M_{J^{''}} = \pm 1$), we report a somewhat lower maximum efficiency than previous groups which we attribute to the aforementioned imperfect laser polarization in our experiments. In addition, we see a deviation of $\eta_{i,\mathrm{max}}$ for optical pumping with $\sigma^+$ and $\sigma^-$-polarized light especially while exciting via the $2^3$P$_1-2^3$S$_1$ transition. This might be induced by a systematic asymmetry in our setup resulting from e.g. small magnetic-field inhomogeneities as discussed above.
\begin{table}[ht!]
	\caption{\label{tab:optPumpEfficiencies} Summary of maximum efficiencies $\eta_{i,\mathrm{max}}$ obtained for the laser optical pumping of He($2^3$S$_1$) into selected $M_{J^{''}}$ sub-levels in our experiment, and comparison with literature values. The given uncertainties (two standard deviations) of our experimental results are statistical only.}
	\begin{threeparttable}
		\begin{tabular}{@{}lcccccc@{}}
			\toprule[0.7pt]
			&& \multicolumn{5}{c}{$\eta_{i,\mathrm{max}}$ (in \%)}\\
			\cmidrule[0.7pt]{3-7}
			&& $M_{J^{''}} = +1$ && $M_{J^{''}} = 0$ && $M_{J^{''}} = -1$ \\
			\midrule[0.7pt]
			$2^3$P$_2-2^3$S$_1$ transition &&  &&  && \\
			This work && $94 \pm 3$ && -- && $90 \pm 3$ \\
			Granitza et al. (1995) \cite{Granitza1995} && $98.5$ && -- && $98.5$ \\
			Lynn et al. (1990) \cite{Lynn1990} && $96$ && -- && $96$ \\
			Giberson et al. (1982) \cite{Giberson1982} && $\approx 66$ && -- && $\approx 66$ \\
			\midrule[0.7pt]
			$2^3$P$_1-2^3$S$_1$ transition &&  &&  && \\
			This work && $87 \pm 5$ && $93 \pm 4$ && $75 \pm 5$ \\
			Granitza et al. (1995) \cite{Granitza1995} && $<98.5$\tnote{a} && -- && $<98.5$\tnote{a} \\
			Wallace et al. (1995) \cite{Wallace1995} && $97$ && -- && $97$ \\
			\midrule[0.7pt]
			$2^3$P$_0-2^3$S$_1$ transition &&  &&  && \\
			Kato et al. (2012) \cite{Kato2012} && $>99$ && -- && $>99$ \\
			Schearer \& Tin  (1990) \cite{Schearer1990a} && -- && -- && $96$ \\
			Giberson et al. (1982) \cite{Giberson1982} && -- && $56$ && -- \\
			\bottomrule[0.7pt]
		\end{tabular}
		\begin{tablenotes}
			\item[a] No specific values given.
		\end{tablenotes}
	\end{threeparttable}
\end{table}

\subsection{\label{sec:magDefl} Magnetic hexapole focusing}
\begin{figure}[hbt!]
	\includegraphics[width=8cm]{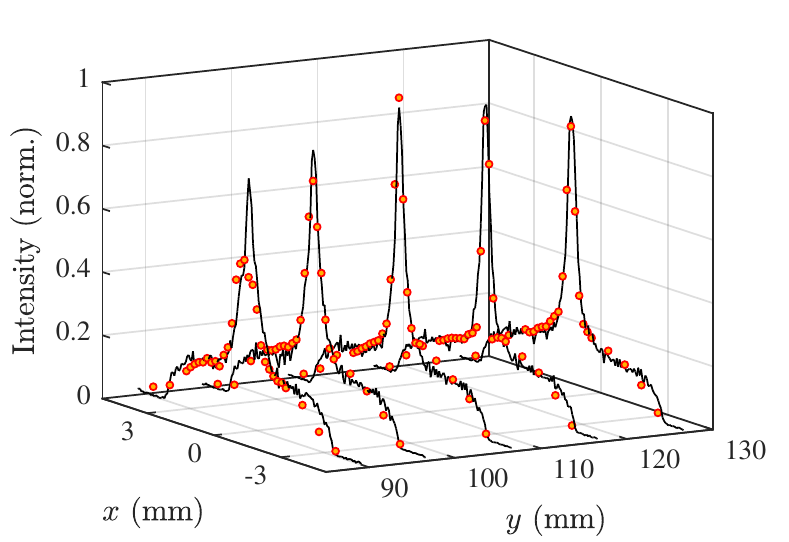}
	\caption{\label{fig:Halbachres} Red circles: Measured He$^*$ signal intensities on the wire detector at different positions $y$ along the supersonic beam axis and at different transverse positions $x$ after magnetic hexapole focusing. The $y$-axis scale is given relative to the center of the two Halbach arrays. Black lines: He$^*$ signal intensities obtained from a numerical particle trajectory simulation. 
	}
\end{figure}

\noindent The red circles in Fig.\ \ref{fig:Halbachres} show the results of a series of measurements which were obtained using the setup for the magnetic hexapole focusing of He($2^3$S$_1$, $M_{J^{''}} = +1$) (cf. Figs.\ \ref{fig:Setup} (c) and (d)). In order to interpret these results, we did numerical three-dimensional particle trajectory simulations in MATLAB. For these simulations, we use random number distributions for the particle positions and velocities (deduced from the experimental data obtained at the Faraday-cup detector) and a velocity-Verlet algorithm. An intial number of $5\cdot10^6$ particles in each Zeeman sub-level of He($2^3$S$_1$) and He($2^1$S$_0$) are propagated at a time. The magnetic field by the two Halbach arrays is implemented using the analytical expressions given in Ref.\ \cite{Dulitz2016}. Particles are removed from the simulation if their transverse position inside a Halbach array exceeds the 3.0 mm inner radius of the assembly (cf. Fig.\ \ref{fig:Setup} (c)).

In each $xy$ detection plane, the output of the trajectory simulation (black lines in Fig.\ \ref{fig:Halbachres}) is analyzed over a certain interval of $x$ positions corresponding to the diameter of the wire detector. The experimental results are matched to the simulated data by comparing the ratio of areas beneath two Gaussian distributions fitted to the datasets (not shown). Very good agreement between the experimental and simulated datasets is achieved by using an effective remanence of $B_{0,\mathrm{eff}} = 1.0 \, \mathrm{T} < B_{0}$ and an effective wire diameter of $d_{\mathrm{wire, eff}} = 5.0 \, \mathrm{mm} > d_{\mathrm{wire}}$ in the simulations. The decreased remanence compared to $B_0$ could be due to the demagnetization of the material as a result of the prolonged storage time of the magnets. Likewise, deviations from the ideal Halbach configuration may also be possible as a result of manufacturing defects.


The analysis of the simulated results suggests that the strong increase of the He$^*$ signal intensity around $x = 0$ is due to the transverse focusing of the $M_{J^{''}} = +1$ sub-level of the $2^3$S$_1$ level. The strongest signal increase, corresponding to the focal point of the device, is at a distance of $\approx 110\,$mm from the center of the two Halbach arrays. The remaining signal intensity is mainly due to a mixture of He atoms in the $2^3$S$_1$, $M_{J^{''}} = 0$ and $2^1$S$_0$, $M_{J^{''}} = 0$ sub-levels. This is also consistent with previous observations \cite{Watanabe2006}. At time $t_0 = 0$, we assume a He($2^1$S$_0$)/He($2^3$S$_1$) ratio of $66\,$\% which is in line with the results of previous measurements in our laboratory \cite{Guan2019}. At the focal point, the signal contribution by He atoms in the $2^3$S$_1$, $M_{J^{''}} = -1$ sub-level is decreased by more than a factor of 7 compared to the signal intensity by atoms in the $2^3$S$_1$, $M_{J^{''}} = 0$ sub-level. This is a result of the strong transverse magnetic defocusing forces which are exerted onto the atoms in the $M_{J^{''}} = -1$ sub-level.

The output of the simulation also provides an estimate of the sub-level selection efficiency for He($2^3$S$_1$, $M_{J^{''}} = +1$), $\eta_{+1}$. Under the conditions of our experiment, $\eta_{+1}$ is nearly constant over a region of $\Delta y \approx 20\,$mm around the focal point. However, the efficiency strongly depends on the He beam diameter considered for the analysis. If we assume that the supersonic beam is collimated to the diameter of the wire detector (i.e., $0.2\,$mm) just in front of this device, we obtain a maximum efficiency $\eta_{+1,\mathrm{max}} = 99\,$\% at the focal point. If we assume the same He beam diameter as in the optical pumping experiments described above (i.e., $2.9\,$mm), the maximum efficiency $\eta_{+1,\mathrm{max}}$ at the focal point is reduced to $84\,$\%. To further improve the sub-level selectivity, we suggest the use of a bent magnetic guide \cite{Beardmore2009, Mazur2014, Osterwalder2015, Dulitz2016, Toscano2018} or by the use of a central stop behind the Halbach arrays  \cite{Chaustowski2007,Kurahashi2008,Kurahashi2021,Baum1988}.

\section{Conclusion}
\noindent We conclude that both laser optical pumping and magnetic hexapole focusing are very efficient methods for the selective preparation of magnetic sub-levels of He(2$^3$S$_1$) in a supersonic beam. We find that optical pumping into the spin-stretched sub-levels of He(2$^3$S$_1$) via the $2^3$P$_2-2^3$S$_1$ transition is more efficient than excitation via the $2^3$P$_1-2^3$S$_1$ transition. The best performance is achieved for $\sigma^{+(-)}$ excitation via the $2^3$P$_2-2^3$S$_1$ transition yielding a maximum efficiency of $94 \pm 3\,$\% ($90 \pm 3\,$\%) for optical pumping into $M_{J^{''}} = +1$ ($M_{J^{''}} = -1$).

Magnetic hexapole focusing is observed to be highly sub-level selective at low forward velocities of the supersonic beam. At $v = 830\,$m/s and at the focal point of the hexapole lens system, we infer that up to $99\,$\% of the metastable atoms are in the $M_{J^{''}} = +1$ sub-level, if an $0.2\,$mm-diameter region around the center of the supersonic beam axis is considered. The magnetic-hexapole-sub-level-selection technique is attractive, because it allows for the quantum-state manipulation of all atomic and molecular species with non-zero spin. Compared to optical pumping, the mechanical setup for magnetic focusing is rather simple, especially when commercial magnets are used \cite{Osterwalder2015}.

However, optical pumping has several advantages compared to magnetic hexapole focusing. While magnetic focusing is limited to the preparation of sub-level-selected samples in low-field-seeking sub-levels only, optical pumping allows for the selective population of all $M_{J^{''}}$ sub-levels, as shown here for the $2^3$P$_1-2^3$S$_1$ transition in He. For optical excitation with $\pi$-polarized light, we obtain an efficiency of $93 \pm 4\,$\% for population transfer into $M_{J^{''}} = 0$. The creation of a pure $M_{J^{''}} = 0$ sub-level might be possible by using magnetic focusing as well but would require a strong overfocusing of the low-field seeking quantum states. In our experiments, this may be realized by further reducing the forward velocity of the He* atoms or by using a longer hexapole magnet array. However, we observe that the number of metastable helium atoms decreases by a factor of $\approx 2$ when the valve temperature is decreased from $300\,$K to $50\,$K. At the same time, the peak He$^*$ flux within the gas pulse decreases by a factor of $\approx 50$, as the longer flight time to the detection region leads to a larger longitudinal spreading of the beam. Optical pumping can be applied independently of the velocity of the atoms as long as the discussed requirements for reaching the equilibrium sub-level efficiency are fulfilled. Thus, this technique results in a greater flexibility in choosing the valve temperature and, as mentioned above, running the valve at higher temperatures leads to much higher peak fluxes of $M_{J^{''}}$-sub-level-selected atoms. These high peak fluxes are particularly important for applications which benefit from high local densities, such as collision experiments. Besides that, optical pumping relies on a transfer of population from a statistical mixture of $M_{J^{''}}$ sub-levels into a single sub-level, whereas magnetic hexapole focusing relies on the spatial focusing (defocusing) of the desired (unwanted) $M_{J^{''}}$-sub-level population. Further transmission losses are due to an aperture which has to be inserted into the beam path behind the magnet assembly in order to eliminate contributions by the $2^3$S$_1$, $M_{J^{''}} = 0$ and $2^1$S$_0$, $M_{J^{''}} = 0$ sub-levels, whose motion is not influenced by a magnetic field.

In the future, we will use the presented sources of $M_{J^{''}}$-sub-level-selected He(2$^3$S$_1$) for quantum-state-controlled Penning-ionization studies \cite{Grzesiak2019}. Furthermore, magnetic-sub-level-selected beams of He(2$^3$S$_1$) are useful as a starting point for the generation of coherent superposition states. The coherent control of Penning and associative ionization cross sections with such superposition states, for instance, involving the $M_{J^{''}} = 0$ sub-level of He(2$^3$S$_1$), has been predicted \cite{Omiste2018}. In addition to that, helium is of particular interest for high-precision tests of few-electron quantum electrodynamics theory, as it is the simplest two-electron atom \cite{Morton2006,Pachucki2017}. Accurate transition frequency measurements have been performed on ultracold trapped samples \cite{vanRooij2011,Notermans2014,Rengelink2018} as well as on atomic beams \cite{Pastor2004,Pastor2012,Zheng2017} of He(2$^3$S$_1$) atoms. The measurement of transitions with zero first-order Zeeman shift (i.e., $M_{J^{'}} = 0 \leftarrow 2^3$S$_1,\, M_{J^{''}} = 0$ transitions) would greatly reduce the experimental uncertainty.

\begin{acknowledgments}
\noindent We thank J. Toscano (JILA) and B. Heazlewood (University of Oxford) for the loan of the magnetic hexapole arrays and for fruitful discussions. This work is supported financially by the German Research Foundation (Project No. DU1804/1-1), by the Fonds der Chemischen Industrie (Liebig Fellowship to K. Dulitz) and by the University of Freiburg (Research Innovation Fund).
\end{acknowledgments}
\appendix
\section{Rate-equation calculations}\label{app:ratEqns}
\noindent The equations used for the characterization of the optical pumping process are of the form
\begin{align}
	\dot{N}^{''}_i &= - N^{''}_i \sum_j W_{ij} + \sum_j W_{ij} N^{'}_j + \Gamma \sum_j \xi_{ij} N^{'}_j, \\
	\dot{N}^{'}_j &= \sum_i W_{ij} N^{''}_i - N^{'}_j \sum_i W_{ij} - \Gamma N^{'}_j \sum_i \xi_{ij},
\end{align}
where $N^{''}_i$ and $N^{'}_j$ denote the populations in the $i$-th and $j$-th magnetic sub-levels of He($2^3$S$_{J^{''} = 1}$) and He($2^3P_{J^{'}}$), respectively, and $\Gamma = 1/\tau$ is the spontaneous decay rate of the excited sub-levels according to their natural lifetime $\tau$. The matrix elements for the excitation rate and for the branching ratio between the $i$-th and $j$-th magnetic sub-level are denoted as $W_{ij}$ and $\xi_{ij}$, respectively. The former are expressed as
\begin{align}
	W_{ij} = \frac{2 \left| \mu_{ij} \right|^2 I}{\hbar^2 \Gamma c \varepsilon_0 (2 J^{'} + 1)} \cdot V(\Delta_{\mathrm{Zeeman}},\Delta_{\mathrm{Doppler}}, \Gamma),
\end{align}
with the laser intensity $I$, the reduced Planck constant $\hbar$, the speed of light $c$ and the vacuum permittivity $\varepsilon_0$. The line-broadening factor $V(\Delta_{\mathrm{Zeeman}},\Delta_{\mathrm{Doppler}}, \Gamma)$ results from a Voigt profile taking into account the Doppler broadening $\Delta_{\mathrm{Doppler}}$, the natural line width $\Gamma$ and the detuning of the transition from resonance due to the Zeeman shift, $\Delta_{\mathrm{Zeeman}}$. The squared dipole matrix element $\left| \mu_{ij} \right|^2$ is calculated using the Wigner-Eckart theorem
\begin{align} \label{eq:dmd}
\left| \mu_{ij} \right|^2 &= \left| \mu_L \right|^2 \cdot \left(2 J^{'} + 1 \right) \left(2 L^{''} + 1 \right) \left(2 J^{''} + 1 \right) \cdot 
\begin{Bmatrix}
L^{''} & L^{'} & 1\\
J^{'} & J^{''} & S
\end{Bmatrix}^2
\cdot
\begin{pmatrix}
J^{'} & 1 & J^{''} \\
M_{J^{'},j} & q & -M_{J^{''},i}
\end{pmatrix}^2,
\end{align}
where $L^{''} = 0$ and $L^{'} = 1$ are the quantum numbers for the orbital angular momenta of the lower and the upper level, $S = 1$ is the quantum number for the total spin and  $q = M_{J^{''},i} - M_{J^{'},j}  = 0,\pm 1$ denote $\pi$ and $\sigma^{\mp}$  polarization, respectively. The spontaneous decay rate $\Gamma$ is used to calculate the squared reduced dipole matrix element $\left| \mu_L \right|^2$:
\begin{align} \label{eq:dme}
	\Gamma &= \frac{\omega^3_0}{3 \pi \varepsilon_0 \hbar c^3} \frac{2 L^{''} + 1}{2 L^{'} + 1} \left| \mu_L \right|^2.
\end{align}
Here, $\omega_0$ is the zero-field transition frequency.

In addition to that, we consider a Gaussian distribution of the laser intensity along the $y$ axis,
\begin{align}
	I = I(y) = I_0 \exp \left(- \frac{2 y^2}{w^2_0} \right),
\end{align}
where $w_0$ is the beam radius and $I_0 = f \cdot 2 P_{\mathrm{laser}}/(\pi w^2_0)$ is the peak intensity calculated from the laser power $P_{\mathrm{laser}}$. The factor $f = 0.1341$ is used to correct for the limited spatial overlap between the laser beam and the supersonic beam. We use the mean forward velocity of the He$^*$ beam in order to transform from the time frame of the rate equations to the position frame of the intensity distribution and to the detector position along the $y$ axis. 

The matrix elements for the branching ratio are calculated using the 3-$j$ symbol
\begin{align}
	\xi_{ij} = (2 J^{'} + 1) \cdot
	\begin{pmatrix}
		J^{''} & 1 & J^{'} \\
		M_{J^{''},i} & -q & -M_{J^{'},j}
	\end{pmatrix}^2.
\end{align}
\section*{Data availability}
The data that support the findings of this study are available from the corresponding author upon reasonable request.
%
%
\providecommand{\latin}[1]{#1}
\makeatletter
\providecommand{\doi}
{\begingroup\let\do\@makeother\dospecials
	\catcode`\{=1 \catcode`\}=2\doi@aux}
\providecommand{\doi@aux}[1]{\endgroup\texttt{#1}}
\makeatother
\providecommand*\mcitethebibliography{\thebibliography}
\csname @ifundefined\endcsname{endmcitethebibliography}
{\let\endmcitethebibliography\endthebibliography}{}

\end{document}